\documentclass[twocolumn]{article}
\usepackage{graphicx}
\usepackage[sorting=none]{biblatex}
\usepackage{amsmath}
\usepackage[margin=1in]{geometry}
\usepackage{xcolor}
\usepackage{booktabs}
\begin{document}

\title{Effective-Hamiltonian Quantum Solvers for Differential Equations: Alternative Constructions and Function Encodings}
\author{Annie E. Paine$^1$}
\date{$^1$ Fujitsu Research of Europe Ltd., Slough SL1 2BE, UK}
\maketitle
\begin{abstract}
	Differential equations can be encoded as ground-state problems by constructing a positive-semidefinite effective Hamiltonian whose minimum-energy state represents the solution. We extend this framework in several directions. First, we show how multiple nonzero initial, boundary, and data conditions can be incorporated into the homogeneous formulation $A | f\rangle=0$ using a known nonzero reference condition. We then analyse an alternative formulation based on $A |f\rangle = | b\rangle$, with Hamiltonian $H_b=A^\dagger(I-|b\rangle\langle b|)A$, which incorporates source terms and nonzero constraints through an augmented system. For nonlinear differential equations, we examine the unphysical ground-state degeneracy introduced by tensor-product representations and consider both Hamiltonian constraints and ansatz-level restrictions for targeting physically valid product states. Finally, we develop grid-value amplitude-encoded versions of both Hamiltonian constructions and compare them with spectral coefficient encoding. Through linear and nonlinear examples, we assess how the choice of Hamiltonian formulation and function representation affects solution recovery, spectral gap, degeneracy, and readout. These results broaden the applicability of ground-state-based quantum differential-equation solvers while clarifying their principal practical trade-offs.
\end{abstract}

\section{Introduction}

Differential equations underpin mathematical models throughout science and engineering, but obtaining accurate solutions can remain computationally demanding, particularly for high-dimensional, nonlinear, stiff, or multiscale problems \cite{han2018solving}. Quantum computers provide an alternative computational paradigm and can offer advantages for selected problems such as certain linear-algebra and simulation tasks \cite{harrow2009quantum,childs2017quantum}. These potential advantages are conditional on factors including data loading, conditioning, target precision, and the information that must be extracted from the resulting quantum state. Nevertheless, the possibility of quantum advantage has motivated a broad range of quantum differential-equation solvers, including quantum linear-system and spectral algorithms, variational and physics-informed methods, and approaches based on linearising nonlinear dynamics \cite{childs2020quantum,liu2021efficient,kyriienko2021solving,paine2023physics}. These methods make different trade-offs, and none provides a universal solution: their applicability or potential advantage may be limited by state-preparation and readout costs, the lack of general trainability and convergence guarantees for variational methods, or restrictions on the classes and parameter regimes of differential equations they can address. Therefore, developing quantum differential-equation algorithms that combine generality, favourable resource requirements, and practical solution readout remains an open problem.

One recently proposed approach encodes a differential equation in an effective Hamiltonian whose ground state represents its solution \cite{wu2025quantum}. In this construction, a function is expanded in a finite set of basis functions, with the expansion coefficients stored in the amplitudes of a normalised quantum state. Wu et al. use Chebyshev polynomials, for which differentiation, multiplication, and function evaluation can be represented by structured operators in the corresponding latent space. After expressing every term of the differential equation in a common basis, these operators are combined into a residual operator $A$ satisfying $A|f\rangle = 0$ if $|f\rangle$ is the exact solution state. Initial, boundary, and data conditions that can be written as homogeneous constraints $B_l|f\rangle=0$ are encoded similarly. The resulting effective Hamiltonian is
$H_{\mathrm{eff}} = A^\dagger A + \sum_l B_l^\dagger B_l$.
This Hamiltonian is positive semidefinite, and any state satisfying all encoded constraints has zero energy. Provided that the common null space uniquely identifies the desired solution, this state is therefore the ground state. For a finite basis or approximate operator implementation, the minimum energy need not be exactly zero, and the ground state instead represents the best solution available within the chosen representation. Once prepared, the Chebyshev-encoded solution can be evaluated at arbitrary points through an overlap with an appropriate feature state.

Several challenges remain within this formulation. First, the homogeneous penalty terms directly accommodate scale-invariant conditions such as $f(x_s)=0$ or $f'(x_s)=0$, whereas general nonzero conditions cannot be inserted independently in the same form because the normalised solution state does not retain the physical scale of the function. Second, polynomial nonlinearities require multiple copies of the coefficient state and an enlarged latent space. The resulting Hamiltonian can possess a degeneracy that grows exponentially with the number of qubits, while most states in this zero-energy subspace need not possess the repeated-product structure required to represent a physical nonlinear solution. Finally, the accuracy, operator complexity, state-preparation cost, and readout procedure all depend strongly on the selected representation.

In this work, we investigate extensions of the effective-Hamiltonian framework that address these issues. We first show how multiple nonzero initial, boundary, and data conditions can be converted into homogeneous, scale-invariant constraints by expressing them relative to a known nonzero reference condition. We then consider an alternative right-hand-side formulation, $A|f\rangle = |b\rangle$, for which the solution is encoded in the ground state of $H_b = A^\dagger \left(I-|b\rangle\langle b|\right)A $. This formulation incorporates source terms and nonzero data through a common augmented linear system. We also investigate additional constraints intended to reduce the unphysical degeneracy introduced by tensor-product representations of nonlinear equations, together with solver-level restrictions that can suppress this degeneracy during ground-state optimisation. Finally, we extend both Hamiltonian constructions to amplitude-encoded grid representations, in which the amplitudes store function values rather than spectral coefficients. Using a range of linear and nonlinear differential-equation examples, we compare the two Hamiltonian constructions and the spectral and amplitude encodings. This provides an analysis of how the choice of representation and Hamiltonian construction affects the applicability of ground-state-based quantum differential-equation solvers. 

\section{Building the effective Hamiltonian}
\subsection{Existing procedure}
We utilise much of the physics-informed effective-Hamiltonian construction of \cite{wu2025quantum} when building the effective Hamiltonian. First a differential equation is considered in zero-residual form 
\begin{align}
    DE(x, f(x), f'(x)...) = \sum_j DE_j(x, f(x), f'(x)...) = 0
\end{align}
where each $DE_j$ is a term of the differential equation. We also define a state $|\psi\rangle$ as representing the function $\psi(x)$ as
\begin{align}
    |\psi\rangle &= (\psi_0, \psi_1,...)^T,~~ \boldsymbol{\phi}(x) =(\phi_0(x), \phi_1(x),...)^T \\
    \psi(x) &= \eta_\psi\sum_j \psi_j \phi_j(x) = \eta_\psi \boldsymbol{\phi}(x)^T |\psi\rangle
\end{align}
where $\{\phi_j\}_j$ is a set of basis functions and $\eta_\psi$ is a scale factor.  For bases admitting an efficient feature map, such as the Chebyshev polynomials used in \cite{wu2025quantum}, a state $\eta_\phi|\phi(x)\rangle = (\phi_0(x), \phi_1(x),...)^T$  can be prepared for arbitrary $x$ within the domain of the basis functions. When this is the case, a function can be evaluated pointwise by measuring the corresponding state overlap as
\begin{align}
    \psi(x) = \eta_\psi \eta_\phi\langle \phi(x)|\psi\rangle.
\end{align}
Otherwise, the amplitudes must be measured to retrieve the coefficients and the basis expansion evaluated classically.

To construct the effective Hamiltonian for a linear differential equation, each residual term is represented by a generally non-unitary latent-space operator $A_j$
\begin{align}
    A_j|\psi\rangle = |DE_j\rangle, ~~ A = \sum_j A_j
\end{align}
where $|\psi\rangle$ denotes a function $\psi(x)$ which we want to approximate the solution $f(x)$ and solve the DE. $|DE_j\rangle$ denotes the coefficient vector of the corresponding residual contribution for $DE_j(x, \psi(x), \psi'(x)...)$ and need not be a normalised quantum state. We note that terms independent of $\psi(x)$ must also be written in this form, this is achieved by utilising a known nonzero reference condition $f^{(m)}(x_s)=f_s\neq0$ and consequently this approach requires that such a condition is known. How to prepare these operators $A_j$ is detailed within \cite{wu2025quantum}. Linear independence of the basis functions then gives
\begin{align}
    A|\psi\rangle = 0,
\end{align}
if and only if $|\psi\rangle$ represents the solution to the differential equation (up to truncation errors). For polynomial nonlinearities, the same construction is performed in an enlarged space acting on multiple copies, $|\Psi\rangle = |\psi\rangle^{\otimes p}$, where $p$ is the maximum polynomial degree.

Homogeneous, scale-invariant initial or boundary constraints, such as
\begin{align}
    f^{(n_l)}(x_l) = 0
\end{align}
are represented by operators $B_l$ such that $B_l |\psi\rangle = 0$ if and only if $\psi^{(n_l)}(x_l)=0$. 

The effective Hamiltonian is consequently
\begin{align}
    H_{\mathrm{eff}}= \mu_{\mathrm{DE}}A^\dagger A+\sum_l \mu_l B_l^\dagger B_l, ~~ \mu_{\mathrm{DE}}, \mu_l > 0
\end{align}
where $\mu_{\mathrm{DE}}$ and $\mu_l$ are optional scale factors which do not alter the exact null space but can affect conditioning, the spectral gap, and the approximate solution when no exact zero-energy state exists. This Hamiltonian is positive semidefinite because $\langle\Psi|H_{\mathrm{eff}}|\Psi\rangle= \mu_{\mathrm{DE}}\lVert A|\Psi\rangle\rVert_2^2+\sum_l\mu_l\lVert B_l|\Psi\rangle\rVert_2^2\geq 0$. The Hamiltonian is constructed such that the energy of a given state corresponds to the error of the associated function in satisfying the given DE and BC constraints. Consequently, if the chosen representation contains a state satisfying all constraints, it is a zero-energy ground state which represents the exact solution. Otherwise, the ground state will correspond to the function in the chosen basis space with lowest DE residual with the ground state energy providing a measure of how well this approximate solution solves the DE. 

Ref.~\cite{wu2025quantum} prepares the low-energy state using quantum imaginary-time evolution implemented through quantum signal processing and singular-value transformation, although other ground-state preparation methods may also be used. Finally, a nonzero reference condition, 
\begin{align}
\label{eq:ref_cond}
    f^{(m)}(x_s)=f_s\neq0,
\end{align}
fixes the remaining overall scale $\eta_{f^*}$ according to 
\begin{align}
    \label{eq:scaleA0}
    \eta_{f^*} = \frac{f_s}{\tilde f^{*(m)}(x_s)}
\end{align}
where $\tilde f^*$ is the unscaled function reconstructed from the prepared ground state. 

With this background, we now introduce and explore the alterations and extensions we propose to generalise this method.

\subsection{Nonzero boundary/initial conditions}
We now consider the inclusion of nonzero boundary, initial, and other data conditions. Let us consider $L$ constraints indexed by $l$ of the form
\begin{align}
    f^{(n_l)}(x_l) = f_l.
\end{align}
These can then be written as
\begin{align}
    f^{(n_l)}(x_l) - f_l = \eta_f \boldsymbol{\phi}(x_l)^T D^{n_l}|f\rangle - f_l = 0,
\end{align}
where $D^{n_l}$ denotes the $n_l^{th}$ degree derivative operator. We can also obtain
\begin{align}
    f_l = \frac{f_l}{f_s}f^{(m)}(x_s) = \frac{f_l}{f_s}\eta_f \boldsymbol{\phi}(x_s)^T D^{m}|f\rangle
\end{align}
utilising the known nonzero reference conditions \eqref{eq:ref_cond} which will be used for scaling and ensure that $f^{(m)}(x_s) = f_s$. Putting these together we have
\begin{align}
    \left(\boldsymbol{\phi}(x_l)^T D^{n_l} - \frac{f_l}{f_s} \boldsymbol{\phi}(x_s)^T D^{m} \right)|f\rangle = B_l |f\rangle = 0
\end{align}

Each resulting homogeneous constraint contributes a positive-semidefinite term to the effective Hamiltonian,
\begin{align}
    H_{\mathrm{IBC}}=\sum_l B_l^\dagger B_l,
\end{align}
so that the full effective Hamiltonian is
\begin{align}
    H_{\mathrm{eff}}=A^\dagger A+\sum_l B_l^\dagger B_l.
\end{align}
A state has zero energy only if it satisfies both the differential equation and all imposed relative data constraints. Consequently, we are now able to impose nonzero conditions as well as zero-valued conditions remaining directly expressible as in \cite{wu2025quantum}. This results in a more generalisable workflow suitable to a wider range of DEs.

\subsection{Alternative effective Hamiltonian form}
Currently the effective Hamiltonian is found by first writing the DE as $A|f\rangle = 0$. An alternative construction can be obtained by adapting the linear-system Hamiltonian used in Ref.~\cite{song2025incompressible}. Consider writing the DE as 
\begin{align}
    DE(x, f(x), ...) = \sum_j DE_j(x, f(x), ...) = r(x).
\end{align}
Here, the left hand side contains any terms of the DE $DE_j(x, f(x), f'(x)...)$ which depend on $f(x)$ whilst the RHS consists of any terms independent of $f(x)$. One can follow the same procedure as previously introduced to construct associated $A_j$ for each term $DE_j$. For the RHS, $r(x)$ is decomposed directly into the chosen basis function representation, and the state which encodes this representation defined as $\eta_b|b\rangle$. This becomes
\begin{align}
    A|f\rangle=\lambda |b\rangle, ~~ \lambda=\frac{\eta_b}{\eta_f}.
\end{align}
The scalar $\lambda$ is not required when constructing the Hamiltonian. Defining the projector onto the subspace orthogonal to $|b\rangle$ as
\begin{align}
    P_b^\perp=I-|b\rangle\langle b|,
\end{align}
the effective Hamiltonian is
\begin{align}
    H_{\mathrm{DE}}=A^\dagger P_b^\perp A=A^\dagger\left(I-|b\rangle\langle b|\right)A.
\end{align}
This operator is positive semidefinite, since
\begin{align}
    \langle f|H_{\mathrm{DE}}|f\rangle=\lVert P_b^\perp A|f\rangle\rVert^2\geq0.
\end{align}
Its energy vanishes precisely when $A|f\rangle \in \mathrm{span}(|b\rangle)$. Provided that the assembled operator $A$ is injective and the finite-dimensional system is consistent, the normalised solution is therefore a zero-energy ground state.  If $A$ has a nontrivial null space, however, states satisfying $A|f\rangle=0$ also have zero energy and may introduce unwanted ground-state degeneracy. The complete operator, including all initial, boundary, and data constraints, must therefore be sufficiently constrained to exclude such null-space states for this type of formulation to be suitable.

Initial and boundary conditions can be included by incorporating them into the same linear system. If 
$C_l\eta_f|f\rangle=d_l$ denotes the l-th condition, one may define
\begin{align}
\tilde A=\begin{pmatrix}A_{\mathrm{DE}}\\C_1\\\vdots\\C_m\end{pmatrix},~~\eta_{\tilde b}|\tilde b \rangle=\begin{pmatrix}\mathbf b_{\mathrm{DE}}\\d_1\\\vdots\\d_m\end{pmatrix}.
\end{align}
The corresponding projector Hamiltonian is then constructed using $\tilde A$ and $|\tilde b \rangle$.

After preparing the ground state $|f^*\rangle$, its physical scale can be recovered from
\begin{align}
    \eta_f=\frac{\eta_{\tilde{b}}\langle f^*|\tilde{A}^\dagger |\tilde{b}\rangle}{\langle f^*|\tilde{A}^\dagger \tilde{A}|f^* \rangle}.
\end{align}
The physical scale could be recovered using the same process as previously \eqref{eq:scaleA0} however this alternate approach means we are not dependent on the existence of a nonzero constraint to provide scaling. This widens the pool of differential equations we can consider. We do note that this method requires $|b\rangle \neq 0$ i.e. at least one DE or data constraint is inhomogeneous.

This formulation naturally accommodates inhomogeneous linear equations and nonzero data. Nonlinear problems still require an enlarged tensor-product representation or an iterative linearisation procedure as previously.

\subsection{Amplitude encoding}
An alternative to representing the solution in a global basis is amplitude encoding, directly encoding function values on a discrete grid with
\begin{align}
    |\psi\rangle = (\psi_0, \psi_1,...)^T,~~\psi(x_j) = \eta_\psi \psi_j
\end{align}
where $\{x_j\}_j$ is a set of grid points and $\eta_\psi$ is a normalisation factor which is determined separately.

To construct an effective Hamiltonian for a given DE within the regime of amplitude encoding, we still need to write our DE as either $A|f\rangle = |b\rangle$ or $A|f\rangle = 0$. Once the DE is in this form, the rest of the procedure follows as previously. 

Consequently, we need to construct the same base blocks which are used to build appropriate $A_j$ for each DE term $DE_j$. The first of these blocks is the derivative operator $D$ which is such that the state $D|\psi\rangle$ will represent the function $\psi'(x)$. To construct this block we utilise a finite difference matrix defined over the same grid $\{x_j\}_j$. A simple example would be a first-order forward-difference operator which implements $f'(x_j) \approx \frac{f(x_{j+1}) - f(x_j)}{h}$ where $h$ is the spacing in the grid $\{x_j\}_j$. Central, backward, or higher-order finite-difference stencils may be used instead, with the choice determining the order of the discretisation error and the treatment of the domain boundaries.

The next block is multiplication by a known function operator $M_a$ which prepares the state $M_a|\psi\rangle$ which represents the function $a(x)f(x)$ where $a(x)$ is a known function which is independent of $f(x)$. As we know $a(x_j)f(x_j) = a(x_j)*f(x_j) ~\forall j$, this operator is able to be simply constructed as $\mathrm{diag}(\{a(x_j)\}_j)$.

We also need to implement nonlinear multiplication operator $N$ such that $|gh\rangle = N|g\rangle|h\rangle$ where $|gh\rangle$ represents the function $g(x)h(x)$ and $g(x)$ and $h(x)$ are arbitrary functions which may depend on $f(x)$. To construct this operator we see that the target state $|gh\rangle$ should be $|gh\rangle = (g_0h_0, g_1h_1,..., g_{2^{N}-1}h_{2^{N}-1})$. We then note that each of these target amplitudes are present within the amplitudes of $|g\rangle|h\rangle$ with $(\langle j | \langle j|)(|g\rangle|h\rangle) = g_jh_j$. With this and generalising to higher orders of nonlinearity, we can form an operator which retrieves and rearranges these amplitudes with $N_D = \sum_{j=0}^{2^N-1} |j\rangle (\langle j|)^{\otimes D}$. 

Finally, to consider data constraints such as initial and boundary conditions, we need an evaluation operator. This is an operator $X_j$ such that $f(x_j) = \eta_fX_j|f\rangle$. For amplitude encoding, $X_j$ must simply select the appropriate amplitude. Therefore, we get $X_j = \langle j|$. With this, we have the building blocks to enable constructing the effective Hamiltonian with amplitude encoding.

Amplitude encoding avoids choosing a truncated global functional basis and is therefore natural when the problem is already formulated on a grid, when finite-difference or finite-volume operators are readily available, or when the solution contains localised or non-smooth features that are poorly represented by global fitting functions. The resulting differential operators are also often sparse and local, though this may not carry forward to $H$.

This flexibility comes at the cost of replacing basis-truncation error with grid and finite-difference error. Smooth functions may require many more grid points than fitting functions such as Chebyshev polynomials, for which spectral convergence can provide an accurate representation using a comparatively small register. Fine-grid differential operators may also become poorly conditioned as $h$ decreases, possibly reducing the spectral gap of the effective Hamiltonian and increasing the importance of preconditioning.

Readout is another important distinction. Recovering the complete grid-value vector requires state reconstruction and at least \(O(2^n)\) classical output. Individual grid amplitudes may instead be queried through computational-basis or interference measurements, although off-grid evaluation requires interpolation. When using spectral encoding, an arbitrary point can be queried directly by preparing the corresponding basis-feature state.

\subsection{Nonlinearity}

We have so far implicitly considered linear DEs but many DEs of interest are nonlinear. In \cite{wu2025quantum} a DE with $D$-degree nonlinearity is considered by utilising $D$ copies of the state of interest $|f\rangle$. E.g. a DE with degree $D$ nonlinearity is written as 
\begin{align}
    A|f\rangle^{\otimes D} = 0 ~\mathrm{or} ~ \eta_b|b\rangle.
\end{align}
The resulting effective Hamiltonian now encodes the solution of the DE within the GS as $D$ copies $|f\rangle^{\otimes D}$. However, with this approach, the solution is not the only ground state, the ground state subspace of $H$ is now highly degenerate. This is because the system is now under-determined. By lifting the space we now have a dimension of $2^{DN}$ but we still only have the constraints from our DE and boundary/initial conditions, which generally does not give us $2^{DN}$ constraints. We now generally have exponentially many ground states but only the ground state of the form $|\psi\rangle^{\otimes D}$ is the correct solution, all the remaining ground states which do not possess this required repeated-product structure do not actually solve the differential equation. This severely limits the usability of this method for nonlinear differential equations. Because of this, we now explore how to overcome this challenge.

First, we consider how to ensure that the outcome of solving the ground state problem is of the form $|f\rangle^{\otimes D}$ by restricting the solver itself. We cosider using a variational approach such as the variational quantum eigensolver (VQE) approach \cite{peruzzo2014variational}. These approaches use a variational ansatz $U(\theta)$ to prepare a trial state $U(\theta)|\psi_0\rangle$. The energy of this state with respect to the Hamiltonian of interest is then minimised by altering the parameters $\theta$. If convergence succeeds, the resulting state shall be the ground state of the Hamiltonian. If we choose a variational ansatz of the form $V(\theta)^{\otimes D}$ and ensure $\theta$ remains the same for both registers then we ensure that all trial solutions shall be of the form $|f\rangle^{\otimes D}$ and so our final solutions shall also be of this form. Hence we have avoided the degeneracy (but not removed its presence within $H$) by restricting our search to states with the required repeated-product structure. With this, variational algorithms which utilise variational ansatz can now target a product-form solution and therefore the solution of the DE.

The approach introduced is restricted to variational algorithms which have their own limitations such as training guarantees and barren plateaus \cite{larocca2025barren}. This raises the question of whether we can reduce the lifted-space degeneracy within the Hamiltonian itself, removing the degeneracy at the fundamental level. A simple approach would be to find a Hamiltonian $H_{sym}$ such that all states of the form $|\psi\rangle^{\otimes D}$ have zero energy and all other states have positive energy, then the sum $H_{DE} + H_{sym}$ will have as its ground state the state which satisfies the DE and is of the correct form, uniquely identifying the solution. Here $H_{sym}$ is providing additional constraints so that $H$ is no longer under-determined. This however is not possible. The ground state space must be a linear subspace and the set of all states of the form $|\psi\rangle^{\otimes D}$ is not a linear subspace e.g. the sum of two such states is not necessarily of the same form $|u\rangle|u\rangle +|v\rangle|v\rangle \neq |w\rangle|w\rangle$. So, we look into what constraints we can add to reduce the degeneracy. We can enforce symmetry. Consider the state $|f\rangle |f\rangle = ( f_0f_0, f_0f_1, ....)^T$. As $f_jf_k = f_k f_j$ we gain a set of constraints of the form $[0,..,0,1,0,...,0,-1,0,...0]|f\rangle|f\rangle = 0$ (generalisable to higher order). This will reduce the degree of the degeneracy however the ground state is still degenerate, more constraints need to be provided. To provide these constraints we will likely require problem specific constraints. For some problems, additional information beyond the DE and the IBCs are known - such as additional data points or symmetries. If enough suitable constraints are known and can be encoded in H, this could break the remaining degeneracy. However, it cannot be guaranteed that a problem of interest will have enough such constraints.

\section{Examples}
We now illustrate the proposed methods using four representative differential-equation problems. Three linear examples (RC-circuit charging, the Airy equation, and steady-state heat conduction through a composite wall) are used to compare spectral coefficient and grid-value amplitude encoding, as well as the homogeneous $A|f\rangle=0$ and projected right-hand-side $A | f\rangle =| b\rangle$ Hamiltonian constructions which we will refer to as $A0$ and $Ab$ type respectively. These examples also demonstrate the treatment of nonzero initial, boundary, and interior data conditions. We then consider the nonlinear Fisher–KPP travelling-front equation. In this case, we restrict the variational search space to repeated-product states, preventing the solver from exploring unphysical directions in the highly degenerate lifted ground-state subspace.

All examples are implemented in Python using OpenQARP \cite{scali2026openqarp}. We employ noiseless statevector simulation, allowing the prepared states and reconstructed solutions to be examined directly. The purpose of these examples is to validate the Hamiltonian constructions and compare their representations, constraints, and solution recovery.

\subsection{Linear}
\subsubsection{RC Charging} 

\begin{figure}[!t]
    \centering
    \includegraphics[width=\linewidth]{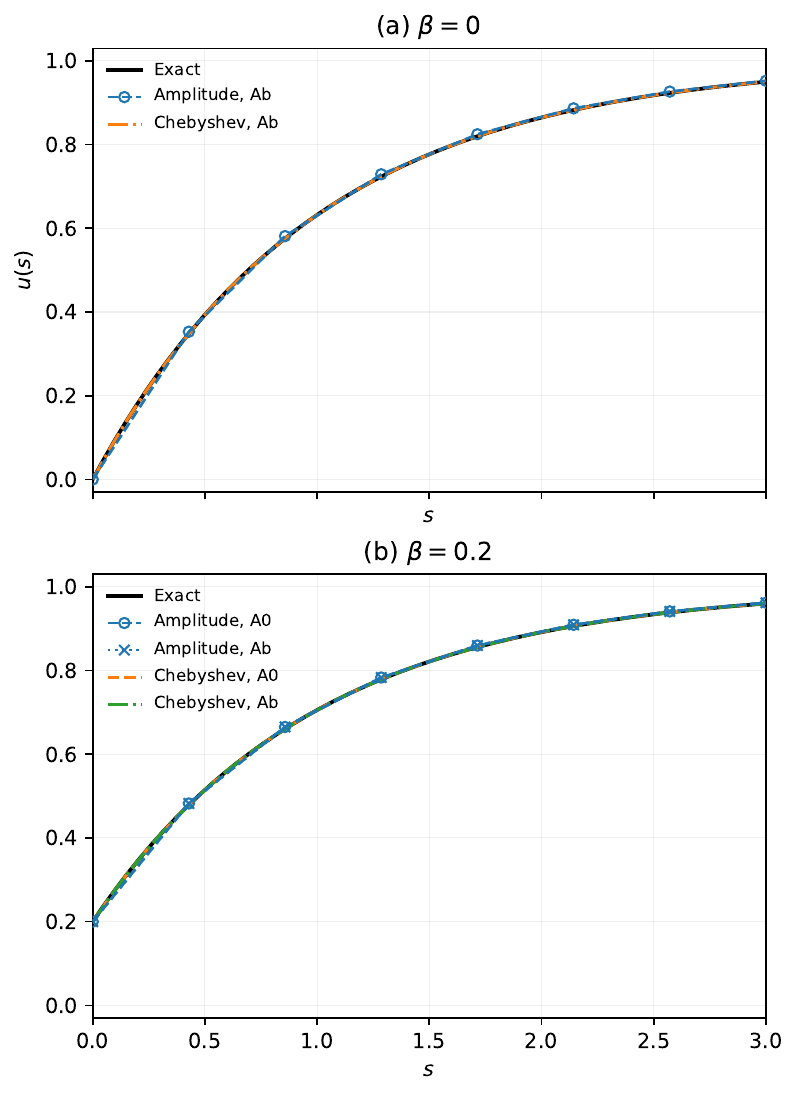}
    \caption{Comparison of the results of solving the RC Charging problem Eq.~\eqref{eq:RC_nondim} to the analytical solution (solid line) using Chebyshev (dashed lines) and amplitude encodings (markers) over $N=3$. (a) Results for $\beta=0$, for which only the $Ab$ construction is applicable. (b) Results for $\beta=0.2$, using both the $A0$ and $Ab$ constructions.}
    \label{fig:RC_res}
\end{figure}

We first consider the charging of a resistor–capacitor (RC) circuit. The charge $q(t)$ on a capacitor of capacitance $C$, connected to a constant voltage source $V_s$ through a resistor of resistance $R$, satisfies \cite{RC1,RC2}
\begin{align}
    R \frac{dq}{dt} + \frac{q}{C} = V_s, ~~ q(0) = q_0.
\end{align}
$q_0$ is the initial charge. Introducing the dimensionless variables $s = t/RC, u = q/CV_s, \beta = q_0/CV_s$ gives
\begin{align}
    \label{eq:RC_nondim}
    u'(s) + u(s) = 1, ~~ u(0) = \beta.
\end{align}
This DE has exact solution $u(s) = 1 + (\beta-1)e^{-s}$. This first-order linear equation therefore provides a simple benchmark containing both a nonzero source term and a tunable initial condition.

We consider $\beta=0$ and $\beta=0.2$ over $s\in[0,3]$. Each problem is solved using both Chebyshev coefficient encoding and grid-value encoding, together with the applicable $A0$ and $Ab$ type Hamiltonian constructions. We use $N=3$ qubits, corresponding to $2^N=8$ Chebyshev coefficients or eight uniformly spaced grid points. For the spectral representation, the Chebyshev basis is mapped onto $[0,3]$.

\begin{table*}[t]
    \centering
    \caption{
        Results for the dimensionless RC-charging equation using $N=3$ qubits. The normalized spectral gap is defined as $\Delta_{\mathrm{norm}}=\Delta/\lVert H\rVert_2$.
        The maximum absolute error $\epsilon_{\infty}$ and mean square error $\epsilon_{\mathrm{MSE}}$ are evaluated at the eight grid points.
    }
    \label{tab:RC_results}
    \begin{tabular}{c l l c c c c}
        \toprule
        $\beta$
        & Encoding
        & Construction
        & $E_0$
        & $\Delta_{\mathrm{norm}}$
        & $\epsilon_{\infty}$
        & $\epsilon_{\mathrm{MSE}}$\\
        \midrule
        $0.2$
        & Amplitude
        & $\mathrm{A0}$
        & 1.47e-15
        & 1.726e-2
        & 4.55e-3
        & 1.08e-5 \\

        $0.2$
        & Amplitude
        & $\mathrm{Ab}$
        & 5.10e-16
        & 3.393e-2
        & 4.55e-3
        & 1.08e-5 \\

        $0.2$
        & Chebyshev
        & $\mathrm{A0}$
        & 9.38e-11
        & 4.190e-3
        & 1.89e-6
        & 1.64e-12\\

        $0.2$
        & Chebyshev
        & $\mathrm{Ab}$
        & 9.38e-11
        & 1.832e-3
        & 1.89e-6
        & 1.64e-12\\
        
        \midrule
        $0$
        & Amplitude
        & $\mathrm{Ab}$
        & -2.10e-29
        & 3.082e-2
        & 5.69e-3
        & 1.69e-5\\

        $0$
        & Chebyshev
        & $\mathrm{Ab}$
        & 1.64e-10
        & 1.677e-3
        & 2.38e-6
        & 2.57e-12 \\
        \bottomrule
    \end{tabular}
\end{table*}

To describe the Hamiltonian constructions, let $D$ denote the derivative operator, $Q$ an operator which averages two adjacent amplitudes, $I$ identity, $\mathbf{g_1}$ the vector which represents $g(x) = 1$, and $X_{s_0}$ such that $f(0) = X_{s_0}|f\rangle$. Let $L$ denote the discrete representation of the operator $d/ds+1$, for the Chebyshev representation $L=D+I$. For the amplitude representation $L=D+Q$ where $D$ approximates the derivative over each pair of adjacent grid points and $Q$ maps the amplitude values to the averages of neighbours. The differential equation is therefore enforced at the interval midpoints, and both $D$ and $Q$ map $2^N$ grid values to $2^N-1$ residual values.

For $\beta\neq0$, the source term can be homogenised using the initial condition, giving
\begin{align}
A_0 =L-\frac{\mathbf{g_1}X_{s_0}}{\beta}, ~~ H_0=A_0^\dagger A_0.
\end{align}
Indeed, $A_0| u\rangle=0$ represents
\begin{align}
Lu-\frac{\mathbf{g_1}}{\beta}u(0)=0,
\end{align}
which reduces to the discrete form of Eq.~\eqref{eq:RC_nondim} when $u(0)=\beta$.

For the $Ab$ type construction, the differential equation and initial condition are combined into the augmented system
\begin{align}
    \tilde{A} = \begin{pmatrix}L \\ X_{s_0}\end{pmatrix}, ~~\tilde{b} = \begin{pmatrix}\mathbf{g_1} \\ \beta\end{pmatrix}, ~~ H =\tilde{A}^\dagger P_{\hat{\tilde{b}}}^\perp \tilde{A},
\end{align}
where $\hat{\tilde{b}}$ is normalised $\tilde{b}$.

When $\beta=0$, the $A0$ construction above is undefined because the initial condition cannot serve as the required nonzero reference condition. The $Ab$ construction remains well defined because the augmented right-hand side is nonzero through $|g_1\rangle$, even though its boundary-condition component vanishes. This illustrates an important distinction between the two constructions. For $\beta=0.2$, both constructions are applicable.

Once the effective Hamiltonian has been constructed, we approximate its ground state using the variational quantum eigensolver VQE \cite{peruzzo2014variational}. We employ a $2N$-layer hardware-efficient ansatz \cite{leone2024practical} comprising $R_y$ rotations and a linear pattern of $CNOT$ entangling gates. This gate set restricts the ansatz to real-valued states, desired as we are considering real problems. The variational parameters are initialised uniformly over $[0,2\pi)$ and optimised using BFGS, with a gradient tolerance of $10^{-8}$ and a maximum of $2000$ iterations. Following optimisation, the VQE state is converted into a function using the corresponding grid or Chebyshev representation and its overall scale recovered as appropriate for the Hamiltonian type considered. In all applicable cases, the reconstructed functions closely approximate the analytical solution, as shown in Fig.~\ref{fig:RC_res}.

The two representations differ primarily in their evaluation and discretisation properties. The Chebyshev representation provides a continuous polynomial approximation that can be evaluated throughout the domain, whereas the grid representation directly provides values only at the selected nodes and requires interpolation between them. For the present $N=3$ discretisation, the amplitude encoded Hamiltonians also have larger operator-norm-normalised spectral gaps than their Chebyshev counterparts. However, the relative gap of the $A0$ and $Ab$ constructions depends on the representation, the $Ab$ gap is larger for the amplitude encoding, whereas the $A0$ gap is larger for the Chebyshev encoding. These instance-specific results demonstrate that the choice of construction can affect the spectrum, but they do not establish a general ordering between the two approaches.

\subsubsection{Airy Function}
We now consider the Airy function \cite{NIST:DLMF}
\begin{align}
    \label{eq:Airy}
    \frac{d^2 u(s)}{ds^2} - s u(s) = 0.
\end{align}
This second-order linear differential equation has two linearly independent solutions, $Ai(s)$ and $Bi(s)$, so that its general solution for real $s$ is
\begin{align}
    u(s) =& c_1 Ai(s) + c_2 Bi(s) \\
    Ai(s) =& \frac{1}{\pi}\int_0^\infty \cos \left(\frac{t^3}{3}+st \right) dt, \\
    Bi(s) =& \frac{1}{\pi}\int_0^\infty \left[\exp \left(\frac{-t^3}{3}+st \right) + \sin \left(\frac{t^3}{3}+st \right) \right]dt.
\end{align}
Both independent solutions are oscillatory for $s<0$, whereas for $s>0$, $Ai(s)$ decays exponentially and $Bi(s)$ grows exponentially. Airy functions arise naturally near classical turning points and, for example, in the time-independent Schrödinger equation for a particle in a linear potential \cite{matin2014solving}.

We solve Eq.~\eqref{eq:Airy} over $s\in[-6,3]$ subject to
\begin{align}
    u(0)&=u_0=Ai(0) =\frac{1}{3^{2/3}\Gamma(2/3)},\\
    u'(0)&=u'_0=Ai'(0) =-\frac{1}{3^{1/3}\Gamma(1/3)}.
\end{align}
These conditions select $u(s)=Ai(s)$ from the two-dimensional solution space. Importantly, both conditions are nonzero. They therefore provide an example that cannot be treated directly by the original construction which would require only one nonzero condition.

We consider both the $A0$ and $Ab$ Hamiltonian constructions using Chebyshev coefficient encoding and grid-value encoding. In each case, $N=4$ qubits are used, corresponding to a state dimension of $d=2^N=16$. For the Chebyshev representation, the basis is mapped onto $[-6,3]$. For the grid representation, the amplitudes encode the function values at 16 uniformly spaced points.

For the grid representation, the second derivative is approximated at the interior nodes using the central-difference rule
\begin{align}
u''(s_j) \approx \frac{u_{j-1}-2u_j+u_{j+1}}{h^2}, ~~ j=1,...,d-2.
\end{align}
For the Chebyshev representation, the corresponding first- and second-derivative operators act directly on the coefficient vector. The lowest $d-2$ coefficients of the differential-equation residual are retained, leaving two degrees of freedom to be fixed by the initial conditions.

Let $D_2$ denote the appropriate second-derivative operator, $D_1$ denote the appropriate first-derivative operator, $M_s$ denote multiplication by the coordinate $s$, and $X_{s_0}$ such that $f(0) = X_{s_0}|f\rangle$. The discretized differential operator is then $L=D_2-M_s$.

For the $A0$ construction, the value condition $u(0)=u_0$ is used to recover the overall normalization. The derivative condition is converted into the homogeneous, scale-invariant constraint
\begin{align}
    B = X_{s_0}D_1 - \frac{u'_0}{u_0}X_{s_0}.
\end{align}
The effective Hamiltonian is therefore
\begin{align}
H_0=L^\dagger L+B^\dagger B.
\end{align}
After its ground state has been obtained, it is rescaled so that the reconstructed function satisfies $u(0)=u_0$.

For the $Ab$ construction, the differential equation and both initial conditions are instead combined into the augmented system
\begin{align}
    \tilde A &= \begin{pmatrix} L\\ X_{s_0}\\ X_{s_0}D_1 \end{pmatrix}, ~~
    \tilde{\mathbf b} = \begin{pmatrix} \mathbf{0}\\ u_0\\ u'_0 \end{pmatrix}, \\
    H_b &= \tilde A^\dagger \left( I-
|\hat{\tilde{\mathbf b}}\rangle \langle\hat{\tilde{\mathbf b}}| \right) \tilde A,
\end{align}
where $\hat{\tilde{b}}$ is normalised $\tilde{b}$.

\begin{figure}
    \centering
    \includegraphics[width=\linewidth]{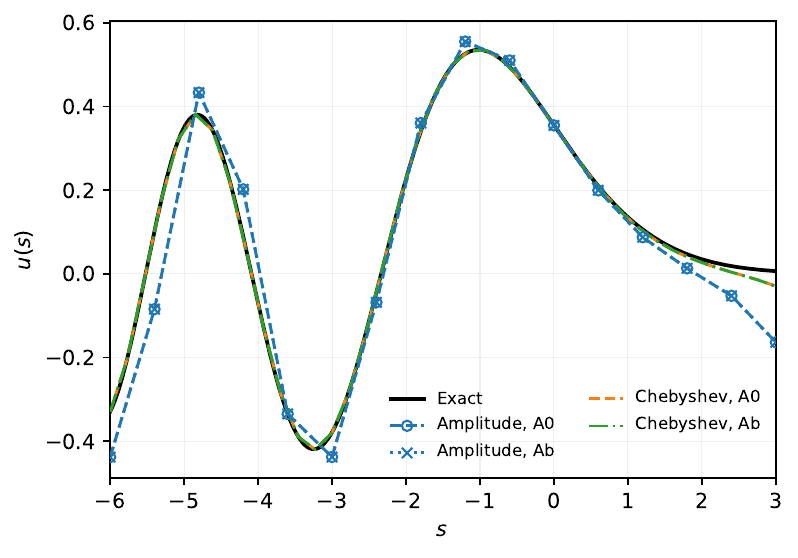}
    \caption{Comparison of the results of solving the Airy equation Eq.~\eqref{eq:Airy} to the analytical solution (solid line) using Chebyshev (dashed lines) and amplitude encodings (markers) over $N=4$ with $A0$ and $Ab$ type Hamiltonians.}
    \label{fig:Airy_res}
\end{figure}

Once the effective Hamiltonian has been constructed, optimisation proceeds the same as for the RC charging example with results shown in Fig. \ref{fig:Airy_res}. We observe that at this fixed number of qubits, the Chebyshev discretization approximates the analytical solution more accurately than the second-order finite-difference discretization for amplitude encoding. The relatively coarse grid spacing $h=0.6$ is insufficient to resolve the oscillatory behaviour accurately throughout the negative-$s$ region. Increasing the number of grid points or employing a higher-order finite-difference rule would reduce this discretization error. The Chebyshev polynomials however are well suited to this oscillatory behaviour. The example therefore illustrates how the choice of function encoding affects the accuracy of the ground state of the resulting Hamiltonian for a given number of qubits.

\subsubsection{Composite Wall}
For our final linear example, we consider steady one-dimensional heat conduction through a composite wall formed from two materials. The wall has total width $L$, with the material interface located at $x=x_I$, and thermal conductivity
\begin{align}
    k(x) = \begin{cases} k_1 ~~0\leq x < x_I \\ k_2 ~~x_I < x \leq L \end{cases}
\end{align}
In the absence of internal heat generation, the temperature $T(x)$ satisfies

\begin{align}
    &-\frac{d}{dx}\left(k(x) \frac{dT}{dx} \right) = 0, \\
    &T(0) = T_h, ~~ T(L) = T_c \\
    &T_-(x_I) = T_+(x_I), ~~ k_1 T_-'(x_I) = k_2 T_+'(x_I)
\end{align}

We nondimensionalise this DE with 
\begin{align}
    s = x/L, u(s) = \frac{T(x)-T_c}{T_h-T_c}
\end{align}

This results in the following DE
\begin{align}
    \label{eq:CompWall}
    &-\frac{d}{ds}\left(k(x) \frac{du}{ds} \right) = 0, \\
    &u(0) = 1, ~~ u(1) = 0 \\
    &u_-(s_I) = u_+(s_I), ~~ k_1 u_-'(s_I) = k_2 u_+'(s_I)
\end{align}
with the solution
\begin{align}
    u(s) = \begin{cases}
        1 - \frac{s/k_1}{R_{tot}}~~ 0 \leq s \leq s_I \\
        1  - \frac{s_I/k_1 + (s-s_I)/k_2}{R_{tot}}~~ s_I \leq s \leq 1
    \end{cases},
\end{align}
with $R_{tot} = s_I/k_1 + (1-s_I)/k_2$.

This is a second order DE with a non-smooth solution. The solution is continuous and piecewise linear, but its derivative is discontinuous at the material interface whenever $k_1\neq k_2$. We use $N=3$, $k_1 = 1, k_2 =10$ and $s_I = 0.4$. For the grid-value representation, the state consists of four uniformly spaced values over $[0,s_I]$ and four uniformly spaced values over $[s_I,1]$. The interface is represented once in each subdomain, and equality of these two amplitudes is imposed through the temperature-continuity condition. The second derivative is approximated within each subdomain using a centred finite-difference stencil, while the interface derivatives are evaluated using second-order one-sided formulas.

We consider two spectral representations. The first uses a single global Chebyshev expansion containing eight coefficients over $[0,1]$. The second uses a domain-decomposed representation containing four Chebyshev coefficients over each of $[0,s_I]$ and $[s_I,1]$. The latter therefore has the same total state dimension as the grid-value and global Chebyshev representations.

For the grid-value and domain-decomposed Chebyshev representations, let the encoded state be written as the direct sum
\begin{align}
|u\rangle= \begin{pmatrix}
    |u_-\rangle\\
    |u_+\rangle
\end{pmatrix}.
\end{align}
Let $D_-^{(n)}$ and $D_+^{(n)}$ denote the $n$th-derivative operators on the left and right subdomains, respectively, and let $X_{s^-}$ and $X_{s^+}$ denote the corresponding function-evaluation rows. For grid-value encoding, these rows select the appropriate grid amplitudes, whereas for Chebyshev encoding they contain the basis functions evaluated at $s_I$. The differential equation and interface conditions are collected into
\begin{align}
A= \begin{pmatrix}
    D_-^{(2)} & 0\\
    0 & D_+^{(2)}\\
    X_{s_I^-} & -X_{s_I^+}\\
    k_1 X_{s_I^-} D_-^{(1)} & -k_2 X_{s_I^+} D_+^{(1)}
\end{pmatrix}.
\end{align}
The first two block rows enforce the differential equation within each material, while the final two rows enforce continuity of temperature and heat flux at the interface.

For the $A0$-type construction, the homogeneous boundary condition $u(1)=0$ is represented by
\begin{align}
B_1= \begin{pmatrix}
    0 & X_{s_1^+}
\end{pmatrix},
\end{align}
giving
\begin{align}
    H_0=A^\dagger A+B_1^\dagger B_1.
\end{align}
The nonzero condition $u(0)=1$ is subsequently used to recover the scale of the ground state.

\begin{figure}
    \centering
    \includegraphics[width=\linewidth]{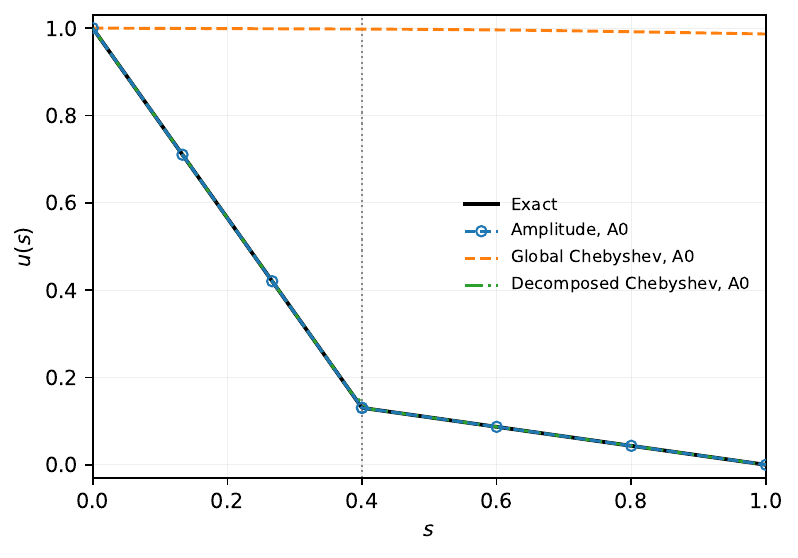}
    \caption{Comparison of the results of solving the Composite Wall DE Eq.~\eqref{eq:CompWall} to the analytical solution (solid line) using Chebyshev and decomposed Chebyshev encodings (dashed lines) and amplitude encodings (markers) over $N=3$ and $A0$ type Hamiltonian. }
    \label{fig:CompWall_res}
\end{figure}

For the $Ab$-type construction, all equations and boundary conditions are instead combined into
\begin{align}
    \tilde A= \begin{pmatrix}
        A\\
        B_0\\
        B_1
    \end{pmatrix}, ~~
    \tilde{\mathbf b}= \begin{pmatrix}
        \mathbf 0\\
        1\\
        0
    \end{pmatrix},
\end{align}
with $B_1= \begin{pmatrix}
    0 & X_{s_0^-}
\end{pmatrix}$. The Hamiltonian is
\begin{align}
    H_b=\tilde A^\dagger \left(I-|\hat{\tilde{\mathbf b}}\rangle\langle\hat{\tilde{\mathbf b}}|\right)\tilde A,
\end{align}
where $\hat{\tilde{\mathbf b}}$ is normalised $\tilde{\mathbf b}$.
For this particular problem, $\tilde{\mathbf b}$ has only one nonzero component. The projector therefore removes precisely the contribution from the $B_0 $row, giving
\begin{align}
H_b=A^\dagger A+B_1^\dagger B_1=H_0.
\end{align}
Thus, the $A0$- and $Ab$-type constructions produce exactly the same Hamiltonian in this example. This equivalence is specific to the present homogeneous equation with one nonzero reference condition and one homogeneous boundary condition; it does not hold for a general nonzero right-hand side. Consequently, only the $A0$-type VQE results are shown in Fig.~\ref{fig:CompWall_res}. Once the effective Hamiltonian has been constructed, optimisation proceeds the same as for the RC charging example.

The grid-value and domain-decomposed Chebyshev results closely reproduce the analytical solution. Both representations can describe this problem particularly efficiently because the exact solution is linear within each material. In contrast, the global Chebyshev construction performs poorly. A single polynomial has the same derivative when the interface is approached from either side. Consequently, the implemented heat-flux condition reduces to
\begin{align}
    (k_1-k_2)u'(\xi)=0,
\end{align}
which incorrectly imposes $u'(\xi)=0$ when $k_1\neq k_2$. The resulting constraint system has no zero-energy state satisfying both boundary conditions, as indicated by its nonzero ground energy. Its poor result should therefore not be interpreted simply as insufficient Chebyshev expressivity: it follows from applying a single smooth representation to a problem requiring distinct one-sided derivatives. Dividing the domain at the material interface resolves this incompatibility while retaining the spectral representation. This example consequently demonstrates the importance of adapting the encoding and domain representation to the regularity of the target solution  as well as the flexibility of amplitude encoding which did not need to be tailored for the example.

\subsection{Nonlinear}

\subsubsection{Fisher-KPP Travelling Front}

We next consider a quadratically nonlinear problem, the travelling-front solution of the Fisher-KPP equation. This reaction-diffusion equation has appications in ecology and takes the form \cite{ablowitz1979explicit}
\begin{align}
    \frac{\partial U}{\partial t} = D \frac{\partial^2 U}{\partial x^2} + rU(1-U),
\end{align}
where $D$ is the diffusion coefficient and $r$ is the linear growth rate. Introducing the transformation $s = x-ct, k = \sqrt{r/6D}$ and $U(x,t) = u(s)$ and considering wave speed $c = 5\sqrt{rD/6}$ gives
\begin{align}
    \label{eq:FKPP}
    u'' + 5ku' + 6k^2u(1-u) = 0,
\end{align}
which has analytical solution
\begin{align}
    u(s) = \left(1 + e^{ks} \right)^{-2}.
\end{align}
Therefore we have a one-dimensional, second-order DE containing a quadratic nonlinearity.

We solve Eq.~\eqref{eq:FKPP} with $k=2$ over $s\in[-1,1]$, using the boundary conditions
\begin{align}
    u(-1)=u_{-1}=\left(1+e^{-k}\right)^{-2},
    ~~
    u(1)=u_1=\left(1+e^{k}\right)^{-2}.
\end{align}
We consider both Chebyshev and amplitude encodings and both the $A0$ and $Ab$ Hamiltonian constructions. The Chebyshev calculations use $N=3$ qubits, whereas the amplitude calculations use $N=4$ qubits to reduce the finite-difference truncation error. This requirement for a larger number of qubits for the amplitude encoding is similar to the Airy example.

To represent the quadratic term, the solution is lifted to the two-register state $|u\rangle\otimes|u\rangle$.
Let
\begin{align}
L=D_2+5kD_1+6k^2I
\end{align}
denote the representation of the linear part of Eq.~\eqref{eq:FKPP} where $D_j$ denotes the $j^{th}$ order derivative operator, and define
\begin{align}
R=\frac{|g_1\rangle X_{s_{-1}}}{u_{-1}},
\end{align}
where $|g_1\rangle$ represents the constant function $g_1(s)=1$ and $X_{s_{-1}}$ such that $f(-1) = X_{s_{-1}}|f\rangle$. Consequently, $R|u\rangle=|g_1\rangle$ whenever the reference condition $u(s_0)=u_0$ is satisfied. Let $N$ denote the product map that converts a tensor-product representation into the representation of the pointwise product. Then we have the DE constraint operator
\begin{align}
    A_\lambda = N\left(R\otimes L-6\lambda k^2 I\otimes I\right),~~ 0\leq\lambda\leq1,
\end{align}
with $\lambda$ controlling the scaling of the nonlinearity, at $\lambda=1$ Eq.~\eqref{eq:FKPP} is represented.

For the $A0$ construction, the two boundary conditions are first combined into the homogeneous constraint
\begin{align}
    B =  X_{s_1}
-\frac{u_1}{u_{-1}} X_{s_{-1}}.
\end{align}
The corresponding boundary Hamiltonian is symmetrised over the two registers,
\begin{align}
    H_{\mathrm{BC}} = \frac{1}{2}\left(I\otimes B^\dagger B+ B^\dagger B\otimes I\right),
\end{align}
and the full Hamiltonian is
\begin{align}
    H_0(\lambda) =A_\lambda^\dagger A_\lambda+H_{\mathrm{BC}}.
\end{align}
The desired nonlinear problem is recovered at $\lambda=1$.

For the $Ab$ construction, the boundary conditions must also be written in the lifted space. We therefore define
\begin{align}
    \langle B_{-1}| &= X_{s_{-1}}\otimes X_{s_{-1}},\\
    \langle B_{1}| &= X_{s_{1}}\otimes X_{s_{1}},\\
    \langle B_{-1,1}| &= \frac{1}{2} \left( X_{s_{-1}}\otimes X_{s_{1}} + X_{s_{1}}\otimes X_{s_{-1}}\right).
\end{align}
The cross-boundary row fixes the relative sign of the two boundary values, which would not be determined by the two squared conditions alone. The augmented system is then
\begin{align}
    \tilde A_\lambda &= \begin{pmatrix}
        A_\lambda\\
        \langle B_{-1}|\\
        \langle B_{1}|\\
        \langle B_{-1,1}|
    \end{pmatrix},
    &
    \tilde{\mathbf b} &= \begin{pmatrix} 
        \mathbf 0\\
        u_{-1}^2\\
        u_1^2\\
        u_{-1}u_{1}
    \end{pmatrix},
\end{align}
giving
\begin{align}
    H_b(\lambda) = \tilde A_\lambda^\dagger \left(I- |\hat{\tilde{\mathbf b}}\rangle
\langle\hat{\tilde{\mathbf b}}| \right)
\tilde A_\lambda,
\end{align}
where $\hat{\tilde{\mathbf b}}$ is normalised $\tilde{\mathbf b}$.

The lifted Hamiltonians act on a $2N$-qubit space and possess highly degenerate ground-state subspaces. For the discretisations used here, numerical diagonalisation gives ground-state degeneracies of $34$ and $47$ for the Chebyshev $A0$ and $Ab$ Hamiltonians respectively, and $211$ and $240$ for the amplitude encoded versions. Most states in these null spaces cannot be written as $|u\rangle\otimes|u\rangle$ and therefore do not represent the square of a single function. The difference in scale between the Chebyshev and amplitude encoding degeneracy is principally due to the difference in number of qubits $N=3$ versus $N=4$.

Rather than searching the complete lifted Hilbert space, we restrict the VQE state to
\begin{align}
    |\Psi(\boldsymbol{\theta})\rangle = |u(\boldsymbol{\theta})\rangle \otimes |u(\boldsymbol{\theta})\rangle,
\end{align}
where the same parametrised circuit and parameters are applied to both registers. This restriction ensures that every state explored by the optimiser has the required repeated-product form. We use the same real-valued hardware-efficient ansatz and BFGS optimisation procedure as in the preceding examples.

The resulting variational objective remains non-convex. Direct optimisation of the complete nonlinear Hamiltonian can converge to alternative low-energy branches of the discretised problem that do not reproduce the desired travelling front. To guide the optimisation towards the branch connected to the analytical solution, we employ numerical continuation. We first solve the linear problem obtained by setting $\lambda=0 $. The resulting parameters are used to initialise the repeated-product ansatz at $\lambda=0.1$. The nonlinear contribution is then increased in increments of $0.1$, with the optimised parameters from each value of $\lambda$ used to initialise the next calculation, until the complete problem at $\lambda=1$ is reached.

\begin{figure}
    \centering
    \includegraphics[width=\linewidth]{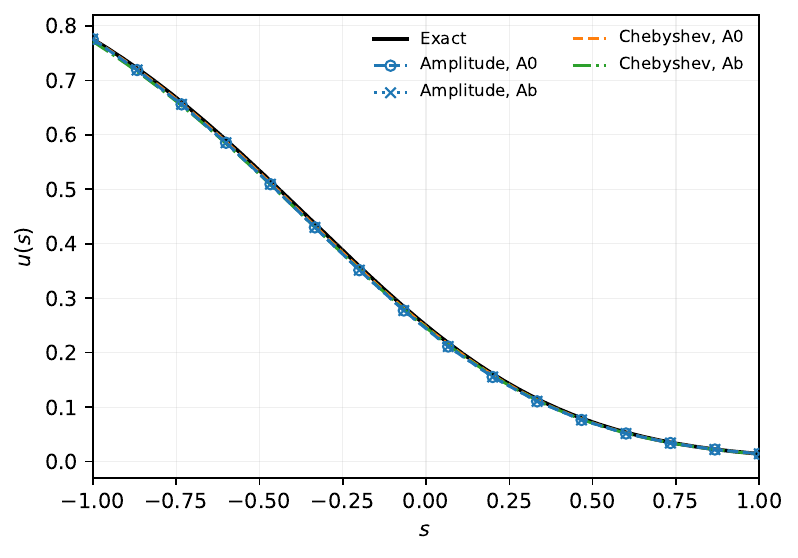}
    \caption{Comparison of the results of solving the Fisher-KPP travelling front Eq.~\eqref{eq:FKPP} to the analytical solution (solid line) using Chebyshev (dashed lines) with $N=3$ and amplitude encodings (markers) with $N=4$. Both $A0$ and $Ab$ type Hamiltonians are considered.}
    \label{fig:FKPP_res}
\end{figure}

The reconstructed solutions are compared with the analytical solution in Fig.~\ref{fig:FKPP_res}. All four combinations reproduce the target solution across the chosen domain. 

\section{Discussion}

In this work, we have extended the effective-Hamiltonian approach to differential equations by incorporating multiple nonzero initial, boundary, and data conditions, introducing the alternative $Ab$ construction, based on $A|f\rangle=|b\rangle$, and extending to amplitude encoding. We have also investigated how repeated-product variational states can be used to treat polynomial nonlinearities in a lifted Hilbert space.

The examples illustrate the effects of these choices. For RC charging with $\beta=0$, the $A0$ type is invalid due to lack of nonzero reference, whereas the $Ab$ construction remains applicable. When $\beta=0.2$, both constructions produce accurate solutions but have different operator-norm-normalised spectral gaps, highlighting that these Hamiltonian types will behave differently. The Airy example shows that both constructions can impose multiple nonzero conditions and that Chebyshev encoding can outperform a finite-difference grid for a smooth solution at fixed resources. Conversely, the derivative discontinuity in the composite-wall solution limits the accuracy of a global Chebyshev expansion. Amplitude encoding performs well without modification, while domain decomposition restores the accuracy of the spectral representation. The Fisher-KPP example demonstrates the treatment of a quadratic nonlinearity. Restricting the VQE ansatz to $|f(\boldsymbol{\theta})\rangle^{\otimes2}$ excludes unphysical non-product states from the search. The full lifted Hamiltonian nevertheless remains degenerate.

Several questions remain open. The present noise-free statevector results demonstrate that the Hamiltonians can be constructed and can encode accurate solutions, but they do not establish efficient quantum ground-state preparation or quantum advantage. The overall cost depends on operator decomposition, state preparation, conditioning, the normalised spectral gap, target precision, and solution readout. Variational implementations additionally face limited expressivity, local minima, barren plateaus, sampling noise, and hardware noise. Such an investigation is still required.

The treatment of nonlinearities also requires further study. The repeated-product ansatz replaces the unphysical degeneracy of the unrestricted search with a constrained, non-convex optimisation problem whose convergence is not guaranteed. Moreover, the present lifting procedure directly represents only polynomial nonlinearities and requires additional registers as the polynomial degree increases. Non-polynomial terms require approximation and introduce further truncation costs. Developing scalable alternatives that enforce the required product structure while retaining reliable ground-state preparation is therefore an important direction for future work.

\printbibliography

\end{document}